\documentstyle[12pt,epsf,epsfig,rotate]{article}

\newcommand \be  {\begin{equation}}
\newcommand \ee  {\end{equation}}
\newcommand \lan {\langle}
\newcommand \ran {\rangle}
\newcommand \sign {\mbox{ sign}}

\begin{document}

\title{Simulation of $3-d$ Ising spin glass model
using three replicas: study of Binder cumulants.}

\author{David I\~niguez$^{(a,b)}$, 
Giorgio Parisi$^{(b)}$ and Juan J. Ruiz-Lorenzo$^{(b)}$\\[0.5em]
$^{(a)}$  {\small  Departamento de F\'{\i}sica Te\'orica, 
Universidad de Zaragoza}\\
{\small   \ \  P. San Francisco s/n. 50009 Zaragoza (Spain)}\\[0.3em]
{\small   \tt david@sol.unizar.es}\\[0.5em]
$^{(b)}$  {\small  Dipartimento di Fisica and Infn, Universit\`a di Roma}
   {\small {\em La Sapienza} }\\
{\small   \ \  P. A. Moro 2, 00185 Roma (Italy)}\\[0.3em]
{\small   \tt parisi@roma1.infn.it  ruiz@chimera.roma1.infn.it}\\[0.5em]}

\date{March, 1996}

\maketitle

\begin{abstract}

We have carried out numerical simulations of the three-dimensional
Ising spin glass model with first neighbour Gaussian couplings using
three replicas for each sample of couplings.  We have paid special
attention to the measure of two types of Binder cumulant that can be
constructed from the three possible overlaps between the replicas.  We
obtain new information about the possible phase transition and perform
an initial analysis of the ultrametricity issue.

\end{abstract}  

\thispagestyle{empty}
\newpage

\section{\protect\label{S_INT}Introduction}

The problem of the existence or not of a phase transition in
3--dimensional spin glasses is still a subject of controversy
\cite{RIEGER}.  In the last year a lot of work has appeared with the
purpose of clarifying the existence or not of the phase transition,
and in the first case, characterize the critical
exponents\cite{KAWAYOUNG,HUKUNEMOTO,MAPARU}.  Moreover recent
numerical work has shown that the low temperature phase is as
predicted by Mean Field \cite{MAPARURI}.  Also recent analytical work
by F.  Guerra \cite{GUERRA} has clarified the meaning of some
formulae found in the framework of the Mean Field Theory.

The most relevant observable in the study of these systems is the
overlap between replicas, {\em i.e.} the order parameter of the
system.  Up to now, all simulations have been carried out with only
two replicas for each sample of couplings. The principal novelty we
introduce in this paper is the use of three replicas.

A way to study a possible phase transition is to use the Binder
cumulant.  This quantity clearly marks the change to a Gaussian
situation from the non Gaussian one.  At present this approach has not
been very successful for spin glasses because at low temperatures the
results turn out to be inconclusive.  One needs a lot of statistics
and large lattices to obtain the small deviation between curves
corresponding to different sizes.

In this paper we study two types of Binder cumulant constructed from
the three different overlaps we can measure between the three
replicas.  The objective is to obtain a clearer signal than in the
case of two replica cumulant.  From one cumulant, which shows a signal
very similar to the two replica ones, we have estimated the critical
temperature and the exponent $\nu$ on the paramagnetic side, obtaining
a good agreement with the previous quoted values \cite{KAWAYOUNG}.
The other cumulant presents new interesting features and supplies us a
method of studying the possible ultrametricity of the spin glass phase
as well as a determination of the critical temperature.

\section{\protect\label{S_SIM} Model, simulation and observables}

The model we have studied is the $3-d$ Ising spin glass with nearest
neighbor couplings distributed Gaussianly around zero. The Hamiltonian
is \be {\cal H}= -\sum_{<i,j>} J_{ij}\sigma_i\sigma_j \ .  \ee As
usual $<i,j>$ denotes nearest neighbor pairs.  The lattice sizes we
simulated are $L=4,6,8,10$.  We have used the simulated tempering
method \cite{TEMPERING,TEMPERING1}.  The range of temperatures studied
has been $[0.7,1.3]$ (for $L=4$ we have also made a run in the range
$[0.3,1.3]$) in steps of $0.05$.  We have simulated 4096 different
samples of couplings for $L=4,6$ and 2048 samples for $L=8,10$, with
three replicas for each sample\footnote{Let be $\sigma, \tau$ and
$\mu$ the three replicas that we will simulate in parallel with the
same disorder.  Hence, we can define three different overlaps
that we
will denote $\{q_{12}, q_{23}, q_{13}\}$ or $\{q, q^\prime, q^{\prime
\prime} \}$ indiscriminately in the rest of the paper.}.

The number of sweeps slightly depends on the lattice size, but it is
typically one million for measuring, after the order of half a million
iterations to estimate the free energy.  The calculations have been
carried out on a {\em tower} of APE100 \cite{APE} with a real
performance, for this problem, of $5$ Gigaflops for a total time of
three weeks.  The errors, sample to sample, have been computed with
the jack-knife method.

As a check of the thermalization procedure we have monitored the
symmetry of the distribution of the overlaps ({\em i.e.}
$P(q)=P(-q)$). We have also checked the two following relations
\cite{GUERRA}:
\be
\overline{ \lan q^2 \ran ^2 } 
= \frac{2}{3} \overline{ \lan q^2 \ran}^2   + 
\frac{1}{3} \overline{ \lan q^4 \ran}
\label{FOR_1}
\ee
\be
\overline{ \lan q^2 {q^{\prime}}^2 \ran} 
= \frac{1}{2}\overline{\lan q^2 \ran}^2 + 
\frac{1}{2}\overline{\lan q^4 \ran}
\label{FOR_2}
\ee
As usual, we denote thermal averages by $\lan (\cdot\cdot) \ran$
and disorder averages by $\overline{ (\cdot\cdot) }$ The first
formula (\ref{FOR_1}) was pointed out in a previous numerical analysis
\cite{MAPARURI}, and have been rigorously demonstrated, along with
(\ref{FOR_2}), by F.  Guerra \cite{GUERRA}.  Both relations are well
satisfied by our data for every value of $L$ and $T$.  Their validity
has been proved in the infinite volume limit, however it is possible
that the deviations are small also for not too large volumes, as those
of the present paper.

We have focused our attention on the following Binder cumulants constructed
from the three overlaps measured for each sample of couplings:
\be
 B_{qqq} \equiv \frac{\overline{\lan \vert q_{12}q_{13}q_{23} \vert \ran}}
        {\overline{\lan q^2 \ran}^{3/2}}
 \ \ \ \ \
 {B^{\prime}}_{qqq} \equiv \frac{\overline{\lan q_{12}q_{13}q_{23} \ran}}
        {\overline{\lan q^2 \ran}^{3/2}}
\ee
and
\be
 B_{q-q} \equiv \frac{\overline{\lan {(\vert q_{12} \vert - 
\vert q_{13} \vert)}^2 \ran}}
        {\overline{\lan {q_{23}}^2 \ran}}
 \ \ \ \ \
  {B^{\prime}}_{q-q} \equiv \frac{\overline{\lan {( q_{12} - 
 q_{13}\sign(q_{23}))}^2 \ran}}
        {\overline{\lan {q_{23}}^2 \ran}}
\ee
where $q_{23}$ is the largest one (in absolute value). These
definitions must follow the finite size scaling relation 
\be
B_{\#} = f_{\#}(L^{1/\nu} (T-T_c))
\ee 
where we have used the symbol \# to denote either cumulant.
In the following sections we will study, in detail, these observables.

\section{\protect\label{S_Bqqq}$B_{qqq}$ and ${B^{\prime}}_{qqq}$}

In Fig.  1 the values obtained from the simulations for $B_{qqq}$ are
shown (lower curves).  We observe that at high temperature the values
for the different lattice sizes are widely separated (the larger $L$
the smaller $B_{qqq}$).  However below $T \approx 1.1$ the curves
intermingle.  In order to distinguish them clearly it would be
necessary to reduce considerably the errors.  The observable
${B^{\prime}}_{qqq}$ presents a similar behavior.  In fact, the
product $q_{12}q_{13}q_{23}$ is generally positive (because it is
favored combinatorially even if one has three completely independent
configurations).

\begin{figure}
\begin{center}
\leavevmode
\centering\epsfig{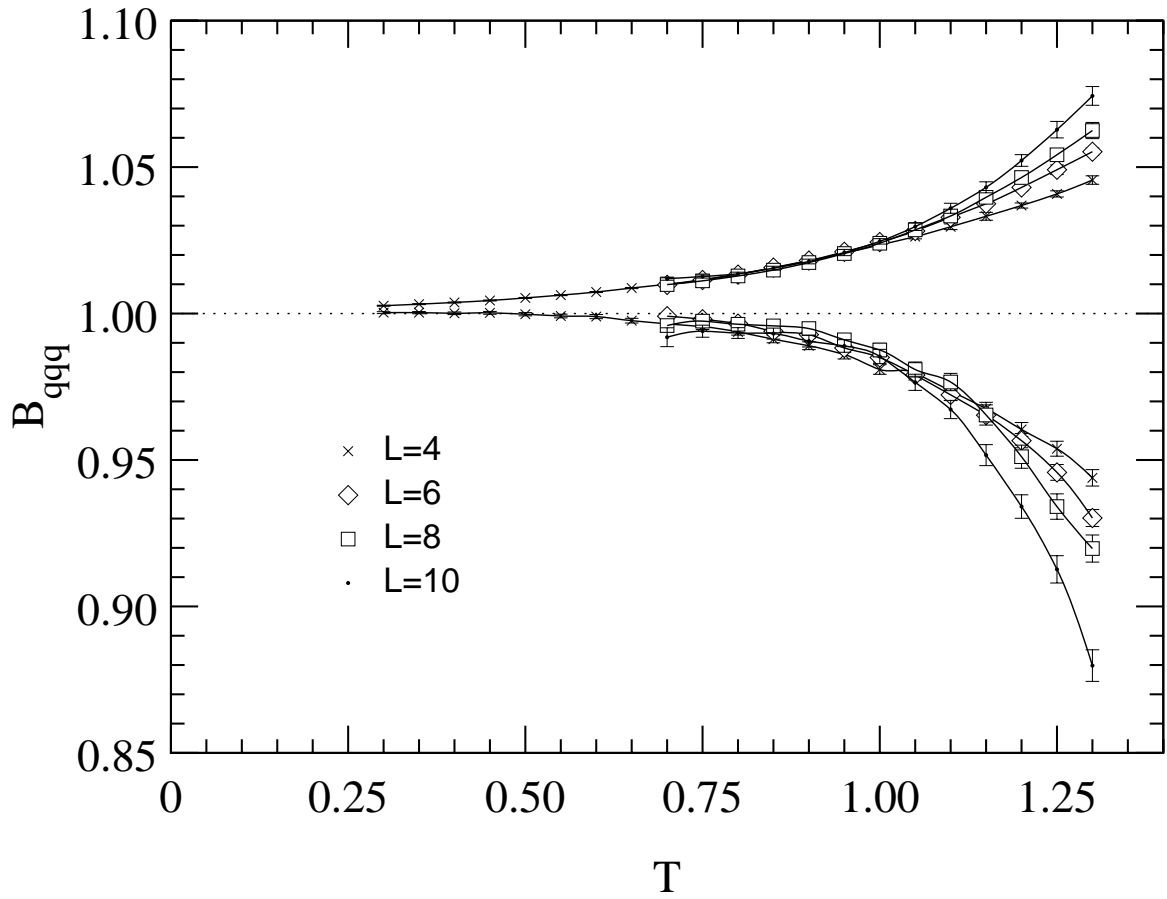}
\end{center}
  \protect\caption[1]{ $B_{qqq}$ against $T$: Numerical data (lower
curves) and supposing ultrametricity (as explained in the text) (upper
curves).
    \protect\label{F_1}
  }
\end{figure}

In the same figure, we show the values we would expect if the three
overlaps satisfied ultrametricity (upper curves).  We have calculated
these curves in the following way.  The probability distribution of
the overlaps for each temperature and lattice size, $P(q)$, is
obtained directly from the simulations.  From it we extract the
function $x(q)$ defined as
\be
 x(q)={\int _0}^q {\rm d}q^{\prime} P(q^{\prime}),
\ee
and then we obtain the inverse $q(x)$, which is shown in Fig.  $2$ for
different temperatures at $L=10$.  Supposing ultrametricity\footnote
{In the appendix we will show that if ultrametricity, convexity and
positivity hold then the functional form of $P_3(q,q^\prime,q^{\prime
\prime})$ must be that of Mean Field for a generic $P(q)$.} \cite{MEPAVI}
$$
P_3(q,q^\prime,q^{\prime\prime})=\frac{1}{2} P(q) x(q)
\delta(q-q^\prime) \delta(q-q^{\prime\prime})+ 
$$
\be
\frac{1}{2}\left( P(q) P(q^\prime) \theta(q^\prime-q^{\prime\prime})
\delta(q^{\prime \prime}-q) +{\mbox {\rm two permutations}} \right)
\ee

the cumulant will take the value
\be
 B_{qqq} = \frac{1}{\left( \int _0^1 {\rm d}x q^2(x) \right)^{3/2}}  
\left(\frac{3}{2} \int _0^1 {\rm d}x q^2(x) \int_x^1 {\rm d}y q(y)
        +\frac{1}{2} \int _0^1 {\rm d}x x q^3(x) \right ) .
\label{binder_ultra}
\ee

We can observe that the curves extracted directly from the simulations
approach those obtained supposing ultrametricity when the temperature
decreases.  This is in agreement with the fact that ultrametricity
could be expected at low temperatures.  However it is not
demonstration because practically any reasonable type of relation
(including independence) between the $q$'s would produce curves
approaching the value $1$ when the temperature goes to zero.  On the
other hand, the curves from the simulations are under the value $1$
(the points for $L=4,\ T=0.3-0.4$ are over $1$ but contain this value
inside their error bars) while the lines bases on ultrametricity
always remain greater than 1.  These results do not rule out the
possibility that the overlaps are as independent as possible for any
temperature (even when the three replicas are totally independent, the
probability of the three overlaps has a constraint given by a simple
combinatorial problem); when decreasing the temperature the behavior
of the $P(q)$ favors the $B_{qqq}$ approaching $1$.

The equation (\ref{binder_ultra}) can also be used to see what would be
obtained if we assume for $q(x)$ a shape that  could be expected in
the thermodynamic limit, that is to say \be q(x)=\cases{q_0
\left(\frac{x}{m}\right)^r,& $0<x<m$, \cr q_0 \hspace*{1.07cm},
&otherwise.  \cr} \ee

We would obtain 
\be
 B_{qqq} = \frac{1}{(1-\frac{2r}{2r+1} m )^{3/2}}
\left[ 1-\frac{3r}{2r+1}m +
\frac{3}{4}\left(1-\frac{2r^2+7r+2}{6r^2+7r+2}\right)m^2
\right].      
\ee

In Fig.  2 we observe that the function $q(x)$ obtained from the
simulations is rather similar to the supposed here with $m=1$ and $r$
descending from approximately 1 for high temperature towards 0 as the
temperature diminishes.  As a test, we can compare qualitatively the
curves shown in Fig.  $1$ in the case of ultrametricity with a
function of this type, and it is seen that in fact the behavior is
similar.

\begin{figure}
\begin{center}
\leavevmode
\centering\epsfig{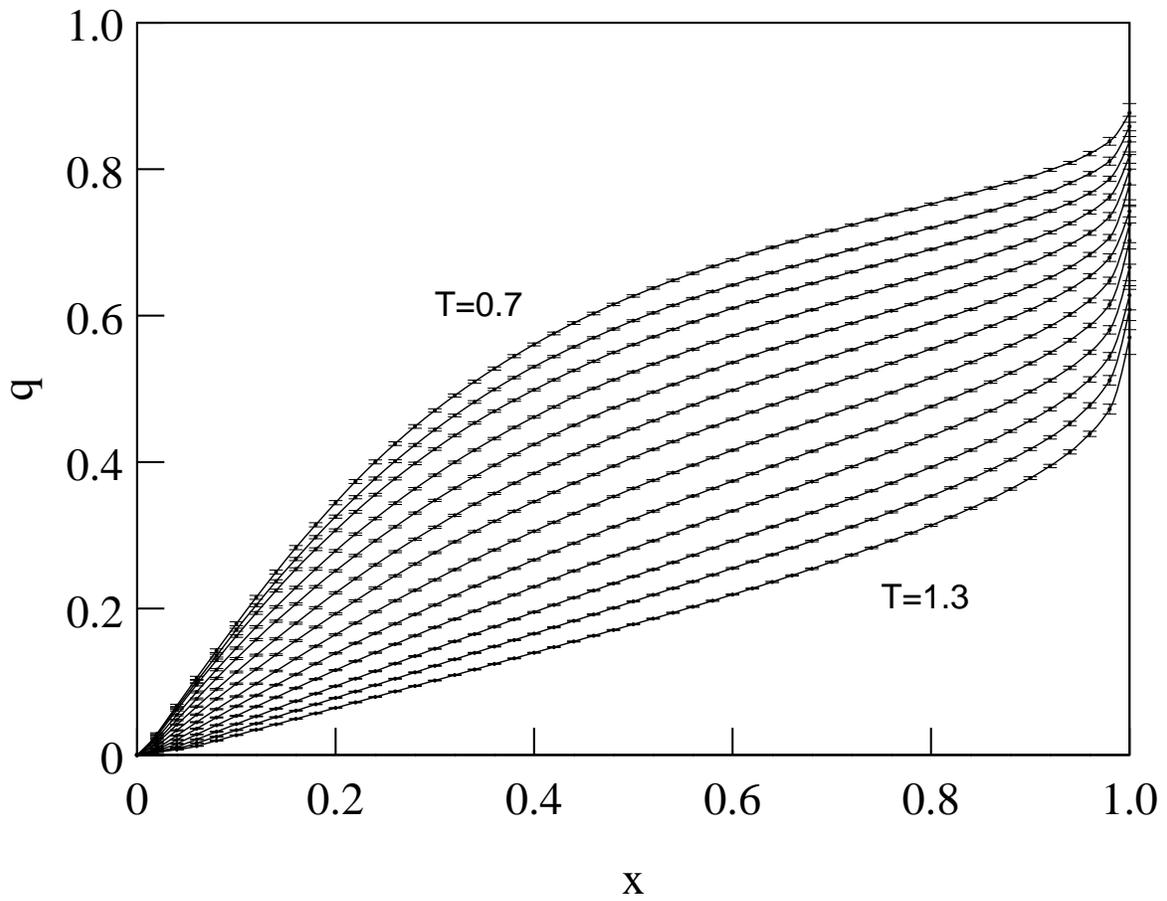}
\end{center}
  \protect\caption[1]{$q(x)$ for 13 different temperatures. 
    \protect\label{F_2}
  }
\end{figure}

We have estimated the value of the exponent $\nu$ by fitting the derivative 
of $B_{qqq}$ in the high temperature region as
\be
f(g,L)\equiv\left. \frac{dB_{qqq}}{dT}\right|_{T_0:\; B_{qqq}(T_0)={\rm g}}
=\alpha
L^{1/\nu} \ .
\ee
We have used polynomials of different orders to fit $B_{qqq}(T)$ and
calculated the derivatives at several values of $B_{qqq}$.  Let us
remark that these derivatives are taken at different values of $T$ but
at fixed $B_{qqq}$.  All the results obtained for $\nu$, for different
values of $g$ near the critical one $g_c\equiv B_{qqq}(T_c)$, are compatible
within the errors, which turn out to be large.  Finally we give an
estimation of $\nu=1.5(3)$, which is in good agreement with the value
$\nu=1.7(3)$ reported in reference \cite{KAWAYOUNG} for the $\pm J$
spin glass.

\begin{figure}
\begin{center}
\leavevmode
\centering\epsfig{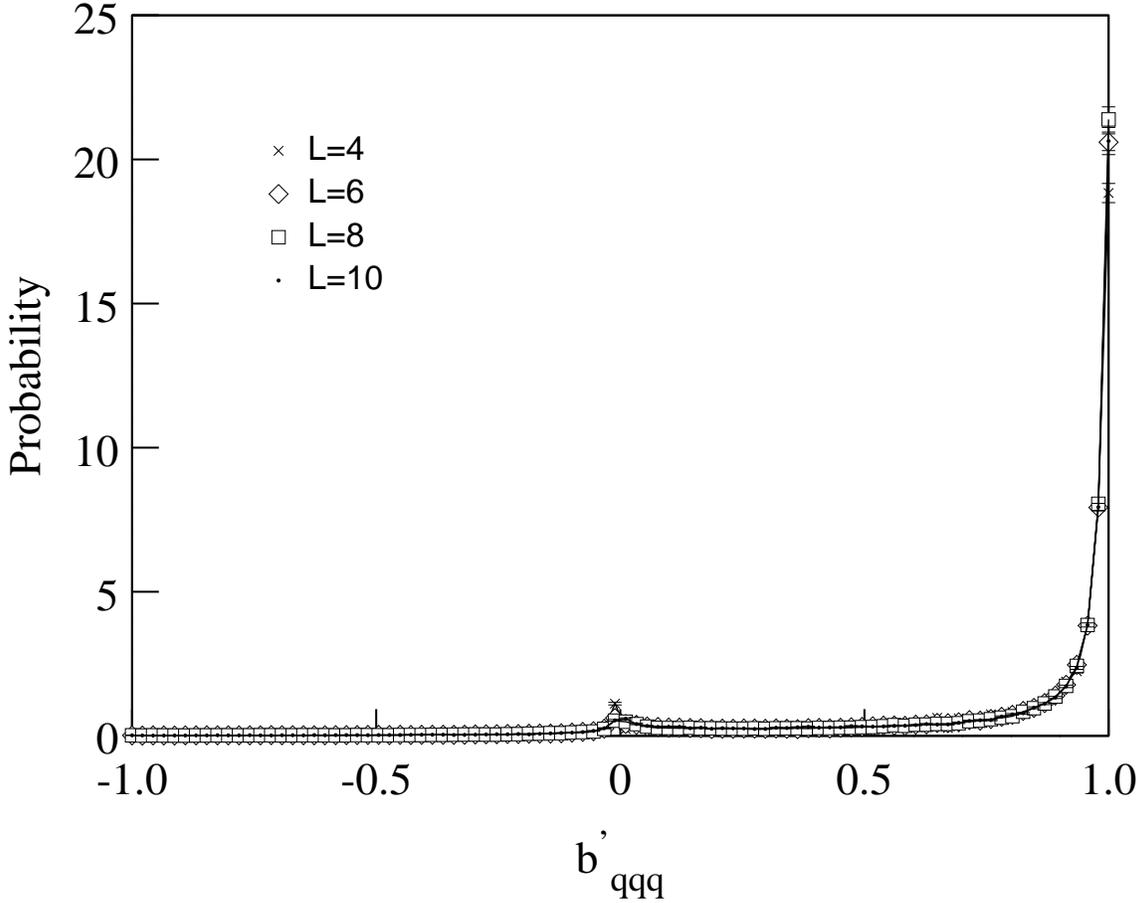}
\end{center}
  \protect\caption[1]{ Probability to have a $b_{qqq}^\prime$ value
for the four different size and $T=0.8$. 
    \protect\label{F_3}
  }
\end{figure}

As we have remarked above, the product $q_{12}q_{13}q_{23}$ is mainly
positive, but not always. In Fig. $3$ we show the probability distribution of
\be
 {b^{\prime}}_{qqq} \equiv \frac{q_{12}q_{13}q_{23}}
        {(\frac{1}{3}(q_{12}^2+q_{13}^2+q_{23}^2))^{3/2}}
\ee
for the different $L$ values at $T=0.8$. Fig. $4$  attempt to show
how the area of the negative part behaves with $L$, but 
it is not clear if it will go to zero with increasing L or not. If mean field 
holds the negative area must go to zero.

\begin{figure}
\begin{center}
\leavevmode
\centering\epsfig{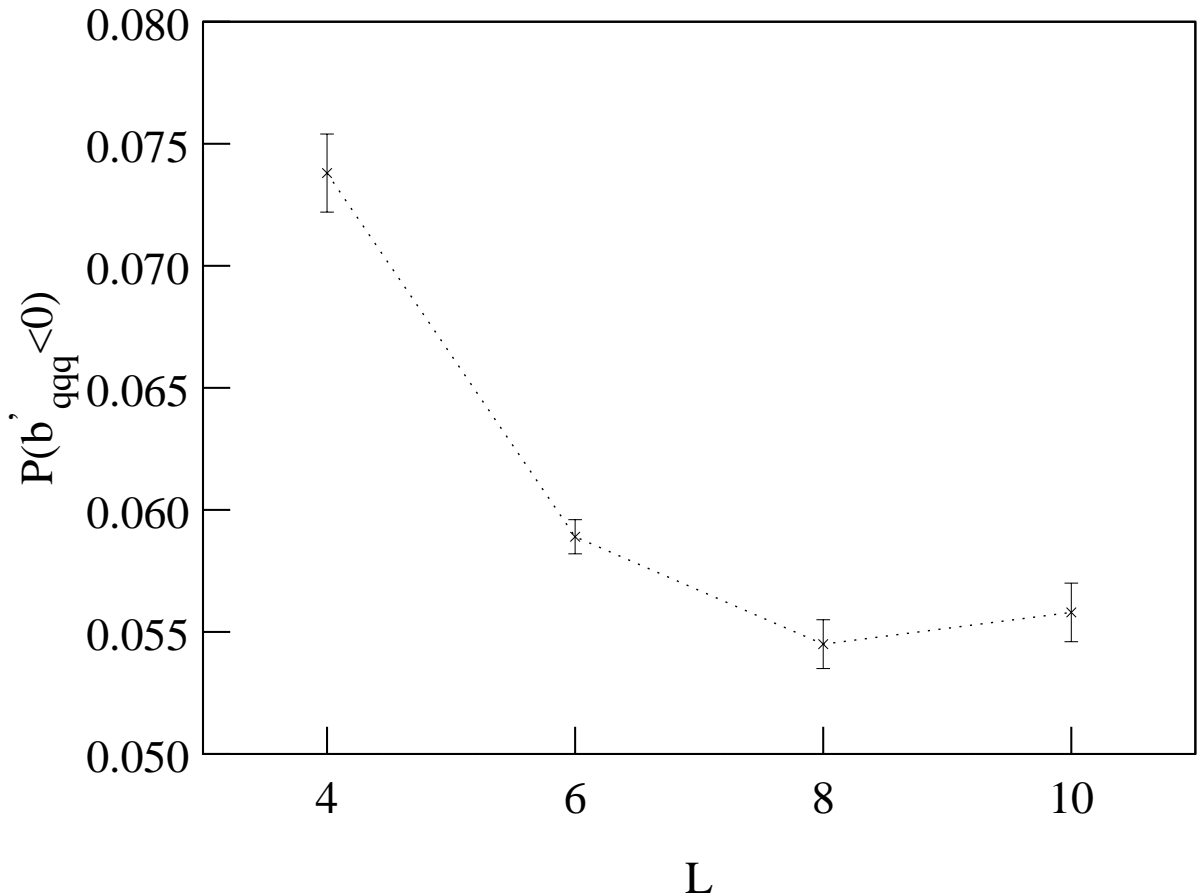}
\end{center}
  \protect\caption[1]{Scaling, as a function of size, of the area of
the negative part ($b_{qqq}^\prime<0$) in figure 3. 
    \protect\label{F_4}
  }
\end{figure}

\section{\protect\label{S_Bq-q} $B_{q-q}$ and ${B^{\prime}}_{q-q}$ }

As can be seen in Fig.  $5$, $B_{q-q}$ has a much clearer signal than
$B_{qqq}$.  We observe a crossing of the curves for different $L$'s in
the region $T\approx 1$.  At low temperatures the values for $L=10$
mask an inversion of the curves (i.e.  the larger $L$ the smaller
$B_{q-q}$ while the high temperature order is the other way round)
which is very clear for $L=4,6,8$, as shown in the expanded Fig.
$5(b)$.  From the crossing of the curves for the larger lattices we
can estimate a critical temperature of approximately $T_c=1.02(5)$.
${B^{\prime}}_{q-q}$ also presents similar behavior.

\begin{figure}
\begin{center}
\leavevmode
\centering\epsfig{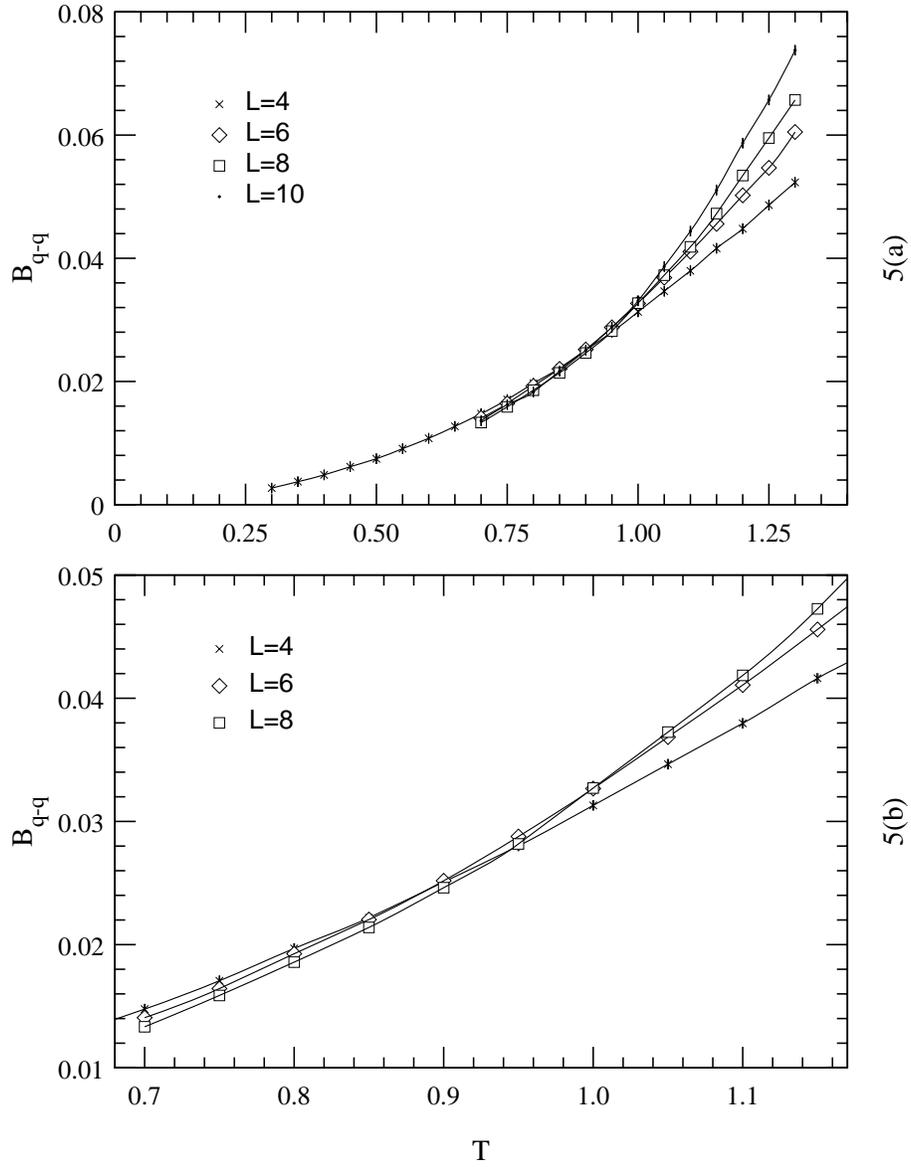}
\end{center}
  \protect\caption[1]{$B_{q-q}$ versus T for the four different
sizes (upper). In the lower part we plot only $L=4,6,8$ and $T\in [0.7,1.3]$. 
    \protect\label{F_5}
  }
\end{figure}

These observables, $B_{q-q}$ and ${B^{\prime}}_{q-q}$, would be zero
if ultrametricity were exactly verified.  In any case, we expect
violations of ultrametricity due to the finite size of our lattice as
happens in the SK model \cite{BATYOUNG}.  The same considerations as in
the preceding section can be done regarding to the approach of the
$B_{q-q}$ curves to zero when the temperature decreases.

\begin{figure}
\begin{center}
\leavevmode
\centering\epsfig{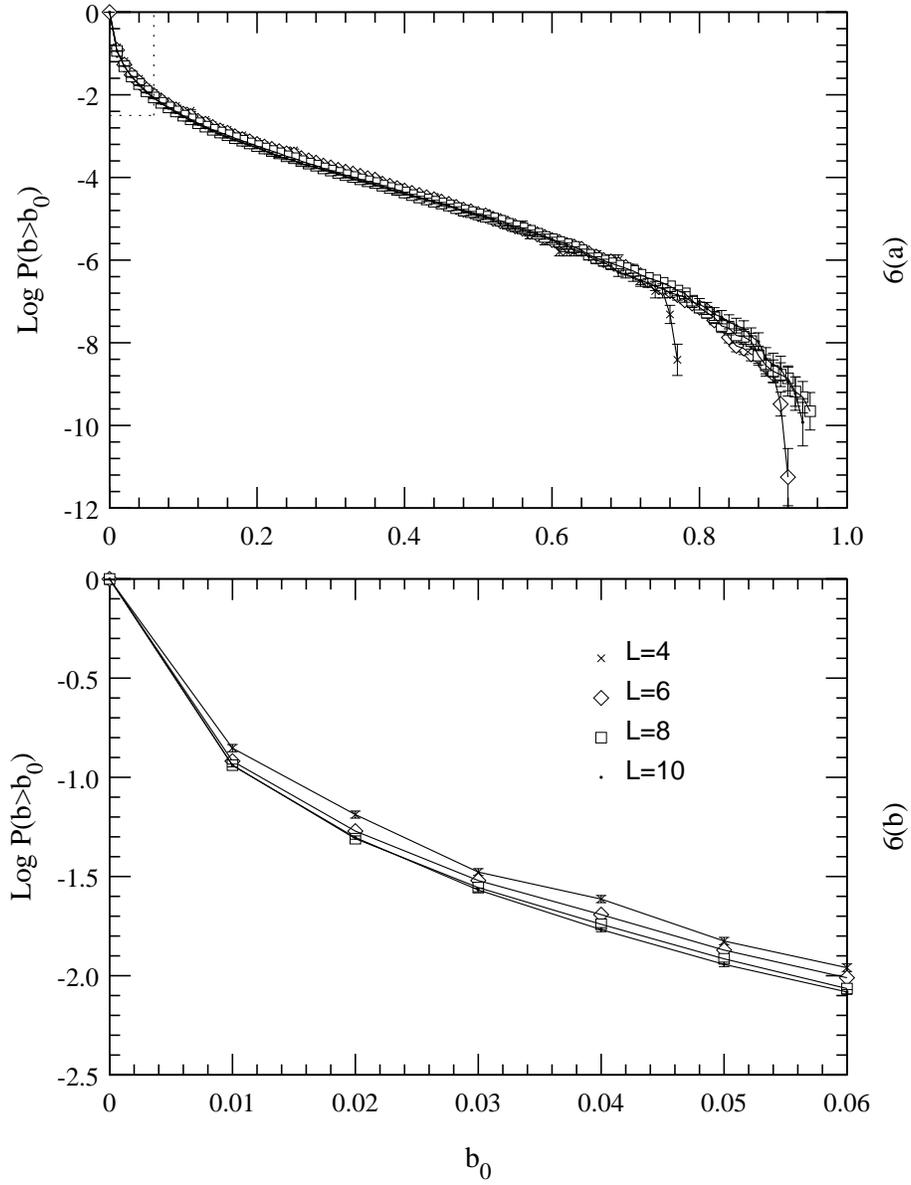}
\end{center}
  \protect\caption[1]{Plot of the accumulated probability for all the
points (upper) and only for a window near the origin (lower).  
    \protect\label{F_6}
  }
\end{figure}

A trial of estimating $\nu$ like above gives results that move
systematically with the value $B_{q-q}$ fixed and then it has been
discarded.

We have made a study of the probability distribution of 
\be
 b \equiv \frac{ {(\vert q_{12} \vert - \vert q_{13} \vert)}^2 }
        {{q_{23}}^2},
\ee
(where $q_{23}$ is the biggest one) calculated at every single iteration,
at temperature $T=0.8$. 

In Fig.  $6$ we show the logarithm of the probability of having a
value of b larger than a certain $b_0$ versus $b_0$.  Fig.  $7$ has
the same variable on the vertical axis but the logarithm of $b_0$ on
the horizontal one.  The curves for the different values of $L$ are
hardly distinguishable.  Expanding the image in the region of small
$b_0$ (figures $6(b)$, $7(b)$), we see that the upper curves are those
of smaller $L$ corresponding to bigger $B_{q-q}$, but the lowering
with $L$ is slow and it is difficult to guess any asymptotic behavior.
Selecting for instance a value of $b_0=0.05$ and observing how this
probability decreases with $L$ we find approximately a power law with
a small exponent of $0.13(2)$.  In the region of large $b_0$, the
upper curves correspond to larger L (in order to maintain the total
probability normalized).  For all the values of $L$, we observe a
first region, for small $b_0$, where the decrease of probability can
be approximated by a power behavior of the form ${b_0}^{-\alpha}$ with
$\alpha=0.71(2)$.  There is a central $plateau$ where it decreases
exponentially as $e^{-\alpha b_0}$ with $\alpha$ ranging from $5.8(1)$
for $L=4$ to $5.3(1)$ for $L=10$, while it goes to zero faster when
$b_0$ approaches the geometrical limit of $1$.

\begin{figure}
\begin{center}
\leavevmode
\centering\epsfig{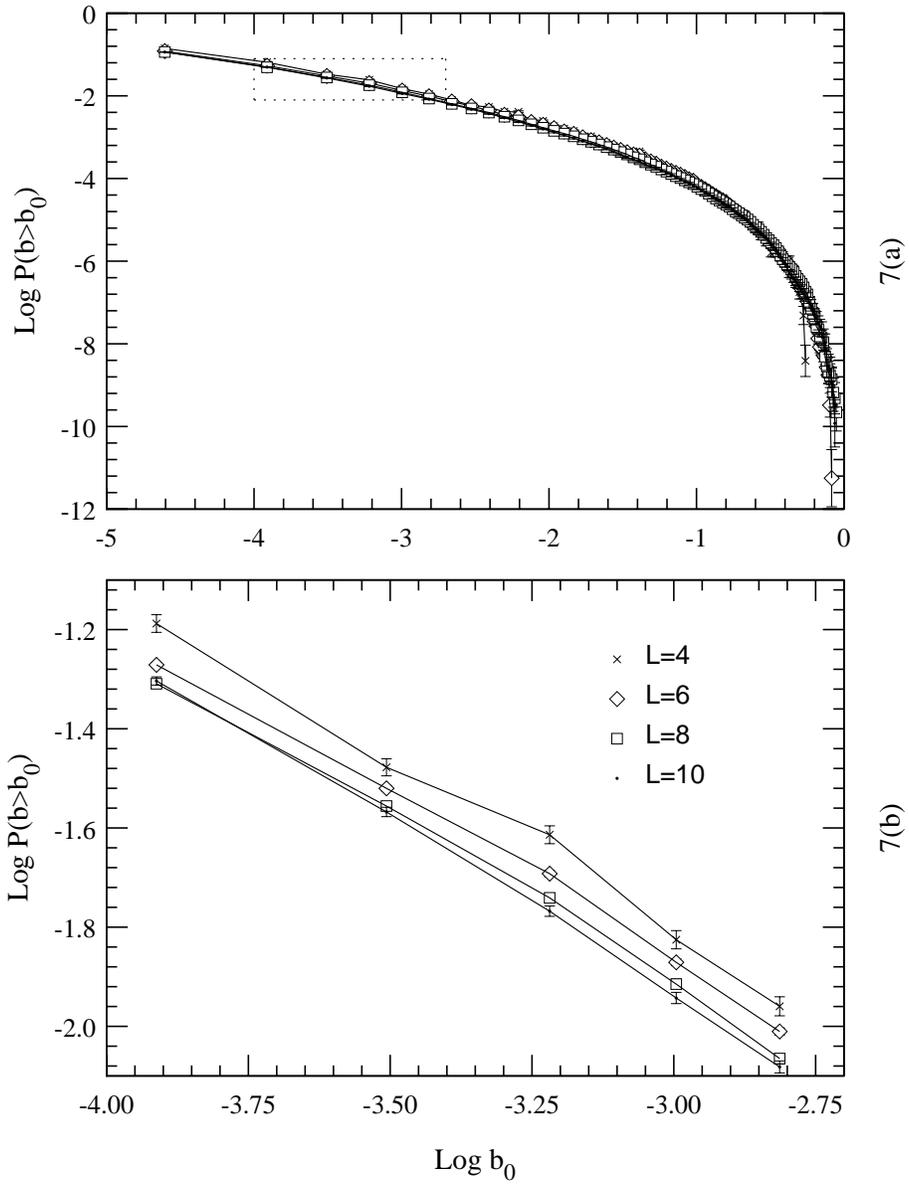}
\end{center}
  \protect\caption[1]{ 
    \protect\label{F_7} The same of Figure 6 but in a double log scale.
  }
\end{figure}

The observable 
\be
 b^{\prime} \equiv \frac{ {( q_{12} - q_{13}\sign(q_{23}))}^2 }
        {{q_{23}}^2}
\ee
shows a similar behavior.

\section{\protect\label{S_CONCLU}Conclusions}

We have reported in this paper a study of the three dimensional Ising
spin glass model using three replicas.  In particular we have studied
different version of the Binder cumulant with three replicas.  We have
estimated both the transition points as well as the critical exponent
$\nu$ being in a good agreement with previous reported values
\cite{KAWAYOUNG}.

Using one of the Binder cumulants we have begun a preliminary study of
ultrametricity.  We have found that the $b$ probability distribution
goes to zero follows a power law. This results are not conclusive (the
exponent is small) and we will need simulate large lattices using new
thermalization methods (for instance parallel tempering
\cite{HUKUNEMOTO}) in order to distinguish between lack of
ultrametricity and possible violations due to finite size effects.

\section{\protect\label{S_ACKNOWLEDGES}Acknowledgments}

We acknowledge useful discussions with E. Marinari and D. Lancaster. 
D.I\~niguez thanks MEC and CAI. J .J. Ruiz-Lorenzo is supported by an EC
HMC(ERBFMBICT950429) grant.

\section{\protect\label{S_APPEN}Appendix}
In this appendix we will show that if ultrametricity holds as well as
convexity and positivity then the three replica probability must be
the Mean Field one.

The most general formula, using the invariance of the Hamiltonian under the 
exchange of replicas, for the probability of three replicas assuming 
ultrametricity is:
 $$ P_3(q,q^\prime,q^{\prime \prime})= A(q)
\delta(q-q^\prime) \delta(q-q^{\prime \prime})+ B(q,q^\prime) 
\theta(q-q^\prime) \delta(q^\prime-q^{\prime \prime}) +
$$
\be
B(q^\prime,q^{\prime\prime}) \theta(q^\prime-q^{\prime\prime}) 
\delta(q^{\prime\prime}-q) +
B(q^{\prime\prime},q) \theta(q^{\prime\prime}-q) \delta(q-q^{\prime}) ,
\ee
with $A(q)$ and $B(q,q^\prime)$ satisfying normalization condition and\-
 $B(x,y)=B(y,x)$. Integrating  $q^{\prime\prime}$ we obtain
the probability to have two replicas
\be
P_2(q,q^{\prime})=\left[ A(q) +\int^{\infty}_q {\rm d}q^{\prime\prime}
B(q^{\prime},q^{\prime\prime})\right ] \delta(q-q^{\prime}) + B(q,q^{\prime}),
\ee 
and finally the probability of one replica ($P(q)$ that we will
denote , in this appendix, as $P_1(q)$) is
\be
P_1(q)=A(q) + \int_q^{\infty} {\rm d}q^{\prime} B(q,q^{\prime}) +
\int_{-\infty}^{\infty} {\rm d}q^{\prime}  B(q,q^{\prime}). 
\label{pq1}
\ee

Now we impose the relation (demonstrated by F. Guerra  using only
positivity and convexity)
\be
\overline{\lan q^2 {q^{\prime}}^2 \ran }\equiv \int_{-\infty}^{\infty} 
{\rm d}q {\rm d}q^\prime P_2(q,q^\prime) q^2
{q^\prime}^2= \frac{1}{2}\overline{\lan q^2 \ran}^2 + 
\frac{1}{2}\overline{\lan q^4 \ran}
\ee
where 
$$
\overline{\langle q^n \rangle} = 
\int_{-\infty}^{\infty} {\rm d} q P_1(q) q^n .
$$

We finally obtain the following equations:
\be
A(q)=\int_{-\infty}^q {\rm d} q^\prime B(q,q^\prime)
\label{pq2}
\ee
\be
B(q,q^\prime)=2 \left ( \int_{-\infty}^{\infty} 
{\rm d} q^{\prime\prime} B(q,q^{\prime\prime}) \right )
\left ( \int_{-\infty}^{\infty} 
{\rm d} q^{\prime\prime} B(q^\prime,q^{\prime\prime}) \right )
\ee
Joining equations (\ref{pq1}) and (\ref{pq2}) 
\be
P_1(q)=2 \int_{-\infty}^{\infty} {\rm d} q^\prime B(q,q^\prime)
\ee
and then we recover the Mean Field formul\ae, {\it i.e.}
\be
A(q)=\frac{1}{2} x(q) P_1(q) ,
\ee
\be
B(q,q^\prime)=\frac{1}{2} P_1(q) P_1(q^\prime)
\ee
with a free $P_1(q)$. Obviously this development does not imply any
particular 
functional form for $P_1(q)$.

\end{document}